\documentclass{pazhb_eng}
\usepackage{graphicx}
\usepackage{natbib}
\usepackage{color}

\usepackage{multirow}

\definecolor{darkblue}{rgb}{0,0,0.9}

\def\smfigure#1#2#3{
  \begin{minipage}{1.0\columnwidth}
    \begin{minipage}{0.049\columnwidth}
      \rotatebox{90}{\small\phantom{0000}#3}
    \end{minipage}
    \begin{minipage}{0.95\columnwidth}
      \includegraphics[bb=40 188 556 678,width=0.97\columnwidth]{#1}
      \centerline{\small #2}
    \end{minipage}

    \vskip 3pt
    ~
  \end{minipage}
}

\def\arcsec{$^{\prime\prime\,}$}
\def\arcmin{$^{\prime\,}$}
\def\fdg{\hbox{$~\!\!^\circ$}}
\def\а{$^{\mbox{\small a}}$}
\def\б{$^{\mbox{\small b}}$}
\def\в{$^{\mbox{\small c}}$}
\def\г{$^{\mbox{\small d}}$}

\def\AA{\buildrel _{\hskip 0.5pt \circ} \over {\mathrm{A}}}

\begin{document}

\journalinfo{2012}{38}{5}{281}[289]

\title{Optical Identification of Four Hard X-ray Sources from the Swift All-Sky Survey}

\author{A.\,A.\,Lutovinov\email{aal@iki.rssi.ru}\address{1}, R.\,A.\,Burenin\address{1}, M.\,G.\,Revnivtsev\address{1},
S.\,Yu.\,Sazonov\address{1}, O.\,N.\,Sholukhova\address{2}, and A.\,F.\,Valeev\address{2}\\
\bigskip
  {\it (1) Space Research Institute, Russian Academy of Sciences, Moscow, Russia}\\
  {\it (2) Special Astrophysical Observatory, Russian Academy of Sciences, Nizhnii Arkhyz, Russia}
}

\shortauthor{Lutovinov et al.}

\shorttitle{Optical identification of four sources}

\submitted{24 December 2011}

\begin{abstract}

We present the results of our optical identifications of four hard X-ray sources from the Swift
all-sky survey. We obtained optical spectra for each of the program objects with the 6-m BTA telescope
(Special Astrophysical Observatory, Russian Academy of Sciences, Nizhnii Arkhyz), which allowed their
nature to be established. Two sources ((\emph{SWIFT\,J2237.2+6324} and \emph{SWIFT\,J2341.0+7645}) are shown
to belong to the class of cataclysmic variables (suspected polars or intermediate polars). The measured
emission line width turns out to be fairly large ($FWHM\sim15-25\AA$), suggesting the presence of extended,
rapidly rotating ($v\simeq400-600$ km/s) accretion disks in the systems. Apart from line broadening, we have
detected a change in the positions of the line centroids for \emph{SWIFT\,J2341.0+7645}, which is most likely
attributable to the orbital motion of the white dwarf in the binary system. The other two program objects
(\emph{SWIFT\,J0003.3+2737} and \emph{SWIFT\,J0113.8+2515}) are extragalactic in origin: the first is a Seyfert 2
galaxy and the second is a blazar at redshift $z=1.594$. Apart from the optical spectra, we provide the
X-ray spectra for all sources in the $0.6-10$ keV energy band obtained from XRT/Swift data.

\englishkeywords{X-ray sources, cataclysmic variables, white dwarfs, active galactic nuclei}

\end{abstract}

\section{Introduction}
\label{sec:intro}

The all-sky surveys currently conducted in the hard ($>15$ keV) X-ray energy band by the INTEGRAL
(Winkler et al. 2003) and Swift (Gehrels et al. 2004) observatories are widely used to solve
various problems: discovering new sources (see, e.g., Krivonos et al. 2007, 2010; Bird et al. 2010;
Baumgartner et al. 2010; Cusumano et al. 2010), investigating the physical properties of various objects
(see, e.g., Filippova et al. 2005; Barlow et al. 2006; Lutovinov and Tsygankov 2009), investigating
the statistical properties of objects of various classes (see, e.g., Sazonov et al. 2007, 2008; Lutovinov et al.
2005; Revnivtsev et al. 2008, etc). The key point for a successful solution of such problems is to
establish the nature of the program objects; the higher the completeness of the identifications, the more
valuable a particular survey and the wider the range of problems it can solve.

To determine the nature of detected hard X-ray sources, various groups, including our group, perform
additional observations with X-ray, optical, and infrared telescopes (see, e.g. Sazonov et al. 2005; Bikmaev et al. 2006, 2008;
Masetti et al. 2007, 2010; Burenin et al. 2008, 2009; Tomsick et al. 2009; Lutovinov et al. 2010, 2012).

In this paper, we determine the nature of four sources from the Swift all-sky survey based on their
X-ray and optical properties. This work is the next one in the program of identifying hard X-ray sources
that we carry out with the BTA and RTT-150 telescopes.

\section{Observations and data reduction}
\label{sec:data}

For our optical observations, we selected four hard X-ray sources (\emph{SWIFT\,J0003.3+2737, SWIFT\,J0113.8+2515, SWIFT\,J2237.2+6324 and SWIFT\,J2341.0+7645}; see the table) from the 58-month Swift catalog (Baumgartner et al. 2010). All these objects were detected with the BAT/Swift
telescope at low flux and confidence ($\sim5\sigma$) levels, which leads to large errors in the localization of
such objects ($\sim6-7$\arcmin, see Tueller et al. 2010). In spite of this, subsequent observations with the Swift
telescope allowed one to identify them in the soft X-ray energy band ($0.6-10$ keV) and to improve
significantly the localization accuracy (to several arcseconds). The latter allowed us to unambiguously
determine the optical object associated with the hard X-ray source.

\bigskip
\begin{table}[]
\centering

\footnotesize{
   \caption{The list of hard X-ray sources}

   \begin{tabular}{lcc}
     \hline
     \hline
          Name &  RA     & Dec     \\
              & (J2000) & (J2000)      \\[1mm]
     \hline
     SWIFT\,J0003.3+2737 &  00$^h$ 03$^m$ 20.2$s$ &  27\fdg 37\arcmin 23\arcsec \\
     SWIFT\,J0113.8+2515 &  01$^h$ 13$^m$ 48.7$s$ &  25\fdg 14\arcmin 38\arcsec \\
     SWIFT\,J2237.2+6324 &  22$^h$ 36$^m$ 26.2$s$ &  63\fdg 27\arcmin 50\arcsec \\
     SWIFT\,J2341.0+7645 &  23$^h$ 40$^m$ 41.8$s$ &  76\fdg 44\arcmin 31\arcsec \\
     \hline
    \end{tabular}
    }
\end{table}

Our optical observations were performed on November 4/5, 2011, with the SCORPIO spectrometer
(Afanasiev and Moiseev 2005) attached to the 6-m BTA telescope (Special Astrophysical Observatory,
Russian Academy of Sciences). We obtained two spectra, each with an exposure time of 600 s, for
\emph{SWIFT J0003.3+2737}; for each of the remaining sources, we took two spectra, each with an exposure
time of 900 s. Here, we present the total spectra for each source, except \emph{SWIFT J2341.0+7645} for which
we detected a shift in the spectra due to the orbital motion of the object in the binary system (for more
details, see below). During these observations, we used the 3D holographic grating \emph{VPHG550G} that
provided the highest light efficiency and a wide spectral range ($3700-7800~\AA$). The spectral resolution
was $\approx6\AA$ (FWHM).

We reduced the optical data in a standard way using the \emph{IRAF}\footnote{http://iraf.noao.edu} software package1 and our own software.
It should be noted that we failed to observe the spectrophotometric standards because of technical
problems. Therefore, below all spectra are presented in relative units and are normalized to the continuum.
The X-ray spectra of the sources in the $0.6-10$ keV energy band were obtained from XRT/Swift data.
They were reduced and analyzed using appropriate software\footnote{http://swift.gsfc.nasa.gov} and the FTOOLS 6.7 software package.

\section{Results}
\label{sec:res}

\bigskip

{\bf SWIFT J0003.3+2737}. The sky field around \emph{SWIFT J0003.3+2727} was observed with the
XRT/Swift telescope on January 21, 2011, with a total exposure time of $\simeq8.4$ ks. Three rather
faint objects are detected in the BAT error circle of the source; one of them is approximately twice as
intense as the other two in the $0.6-10$ keV energy band. Moreover, the last two sources essentially
disappear in the image of the sky field obtained in the $6-10$ keV energy band, while the first object
with (J2000) coordinates RA=00$^h$ 03$^m$ 27.4$^s$, Dec=27\fdg 39\arcmin 18\arcsec (the localization accuracy is $\sim5$\arcsec) is still
detected with confidence. Thus, we may conclude that precisely this object is the soft X-ray counterpart
of \emph{SWIFT J0003.3+2727}.

\begin{figure*}
\centering

\vbox{
~\hspace{5mm}\includegraphics[width=7.0cm,bb=40 140 575 635]{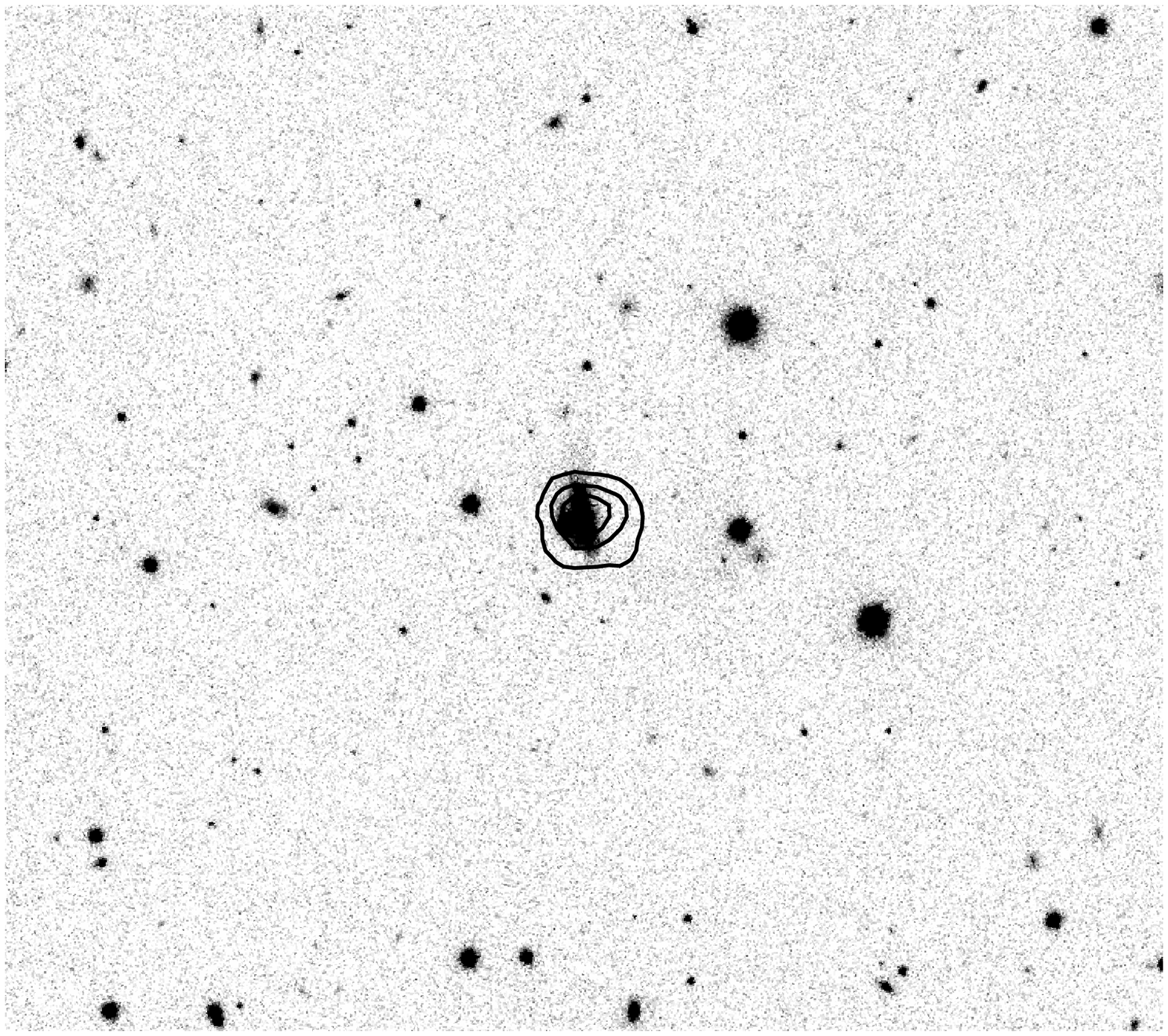}

\hspace{0mm}\smfigure{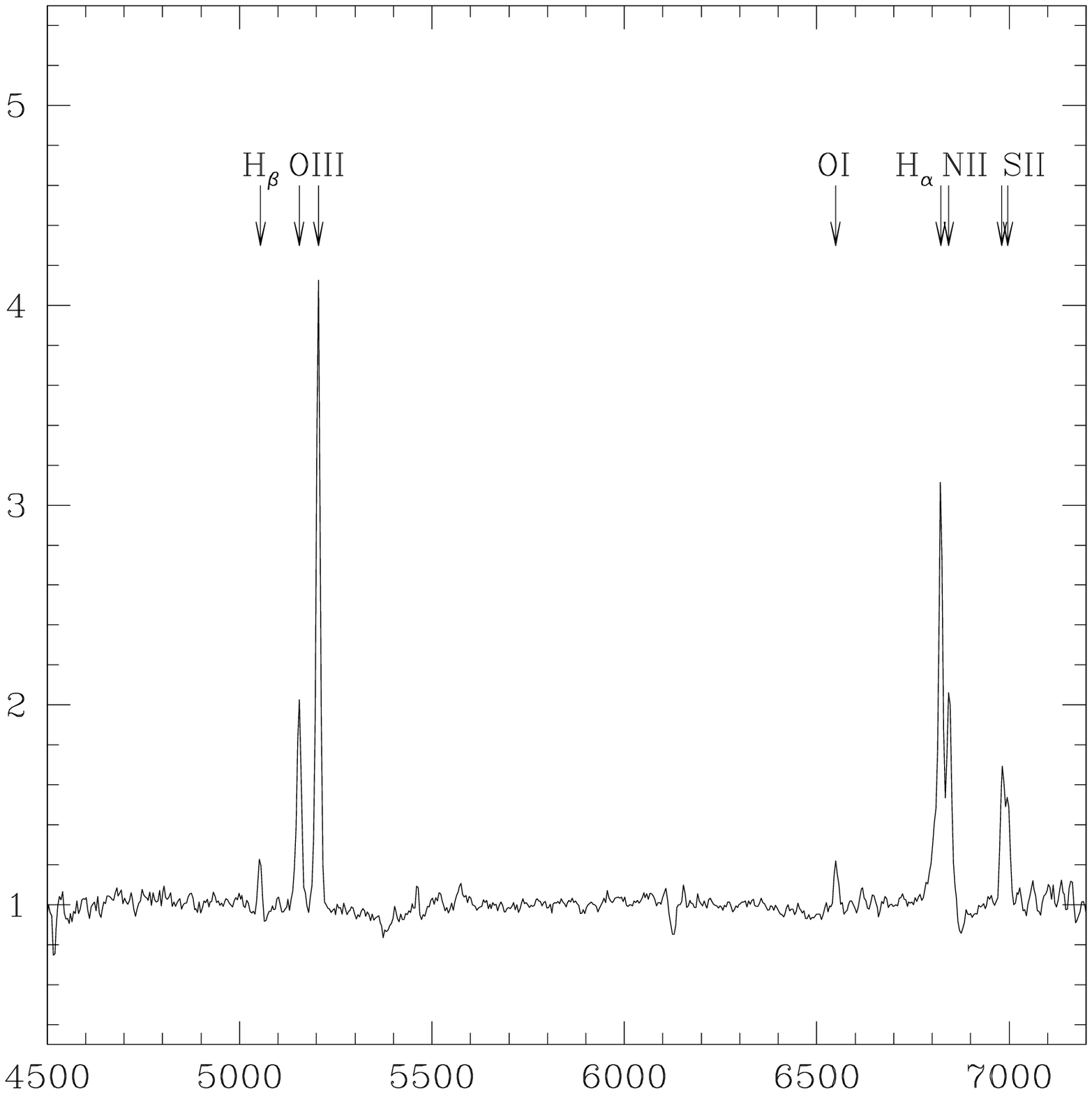}{\small $\lambda$, $\AA$}{\small $F_\lambda$, relative intensity}

\hspace{-2mm}\includegraphics[width=7.9cm,bb=15 165 565 695]{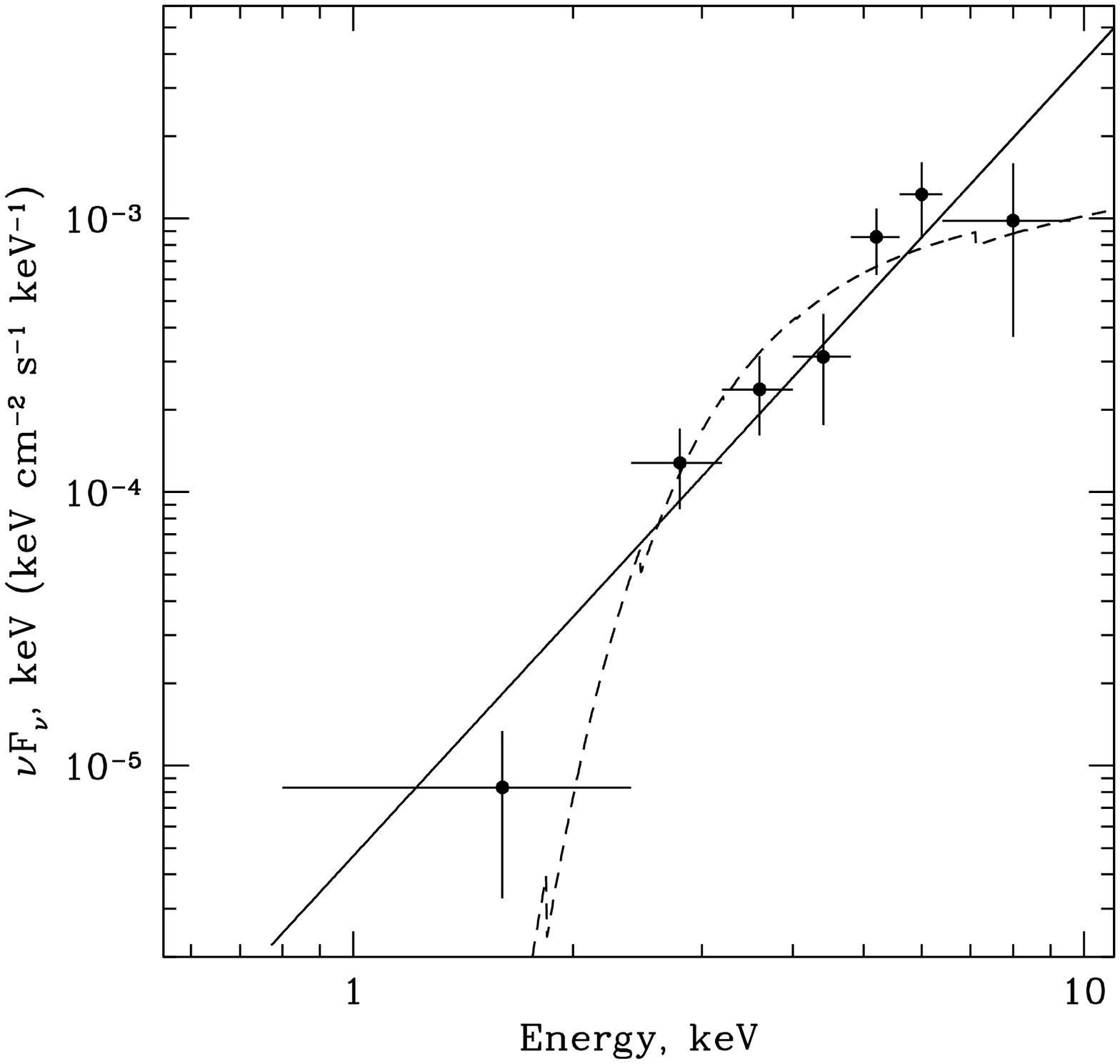}
}

\caption{\small (a) SDSS image of the sky field with \emph{SWIFT J0003.3+2727}; the contours indicate the soft
X-ray intensity levels from XRT data. (b) The optical spectrum of \emph{SWIFT J0003.3+2727} taken with the BTA
telescope. The main emission lines are labeled. (c) The energy spectrum of \emph{SWIFT J0003.3+2727} from XRT
data: the solid and dashed lines indicate, respectively, the best fits by a power law and a power law with a slope of
1.7 and absorption.}\label{swift00033}

\end{figure*}

Comparison of the XRT image with the SDSS images shows that \emph{SWIFT J0003.3+2727} can be
unambiguously identified in the optical band with a relatively bright object, $m_r=16.3$ (Fig.\ref{swift00033}a). Its optical
spectrum clearly shows Balmer emission lines, narrow NII lines, the [OIII] 4959,5007 lines, and
others (Fig.\ref{swift00033}b), typical of a Seyfert 2 galaxy at redshift $z=0.03946\pm0.00002$.

The X-ray spectrum of \emph{SWIFT J0003.3+2727} shown in Fig.\ref{swift00033}c can formally be well fitted by a
simple power-law dependence of the photon flux density on energy $dN/dE\propto E^{-\Gamma}$ with a photon index $\Gamma\simeq-1$.
However, such a photon index is atypical of the spectra of active galactic nuclei (AGNs), that most
likely connected with a significant absorption in its spectrum. Unfortunately, the low quality of the spectrum does
not allow this question to be investigated in detail. We can only fix the photon index at $\Gamma=1.7$, typical
of AGNs and estimate it as $N_H\simeq10^{23}$ cm$^{-2}$ by adding absorption to the model.

The X-ray flux from the source in the $2-10$ keV energy band is $F_{X}\simeq1.5\times10^{-12}$ erg cm$^{-2}$ s$^{-1}$,
which corresponds to a luminosity $L_{X}\simeq5\times10^{42}$ erg s$^{-1}$ for the measured $z$.

\bigskip

{\bf SWIFT J0113.8+2515.} The source \emph{SWIFT J0113.8+2515} was observed with the XRT telescope
three times with a time difference of about one year (February 28, 2010; March 1 and 2, 2011), the total
exposure time being $\simeq8$ ks. Since we found no significant changes in the detected flux and the shape
of its spectrum, we analyzed the total X-ray emission from the program object.

\begin{figure*}
\centering
\vbox{

~\hspace{5mm}\includegraphics[width=7.0cm,bb=40 140 575 635]{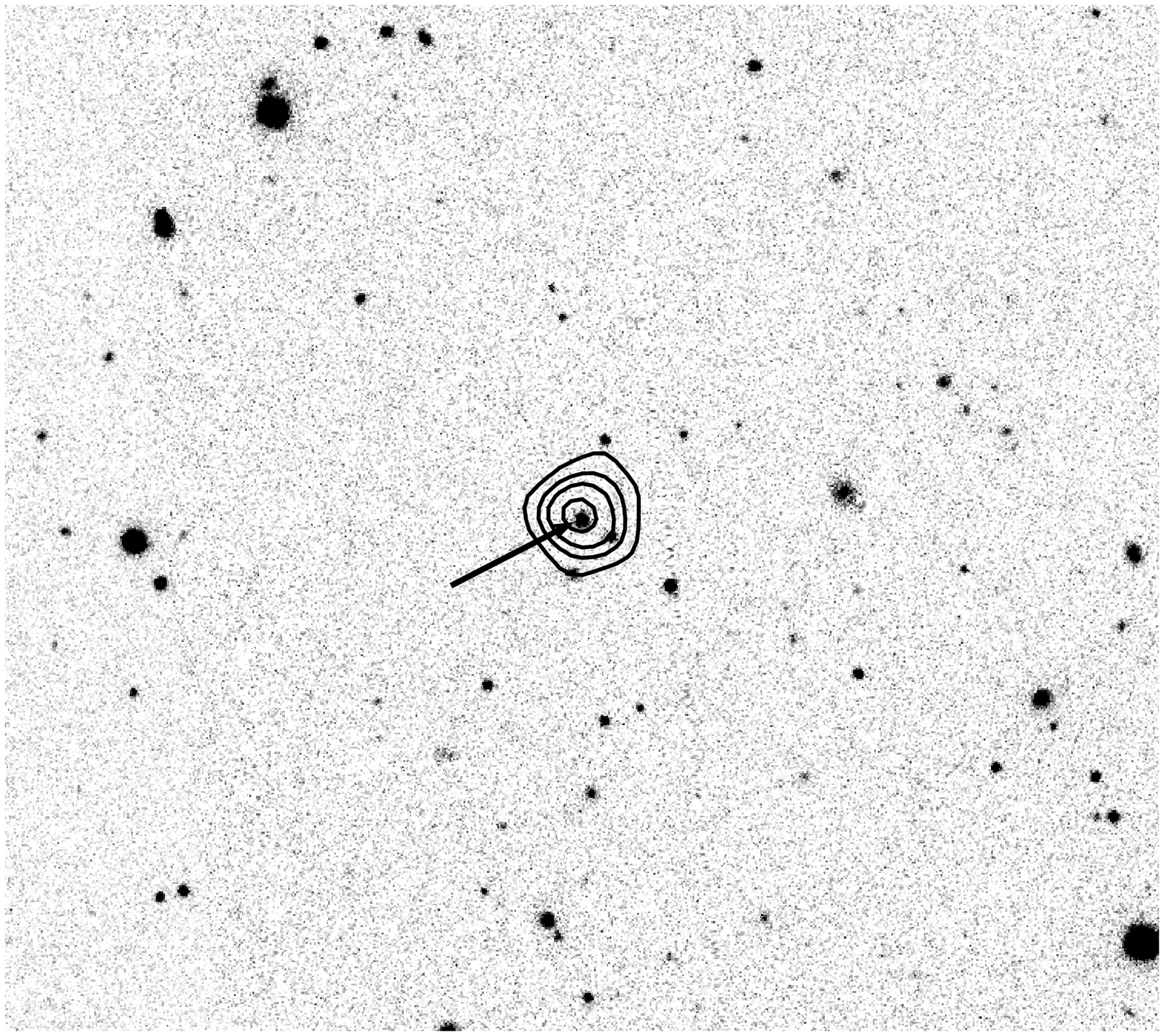}

\smfigure{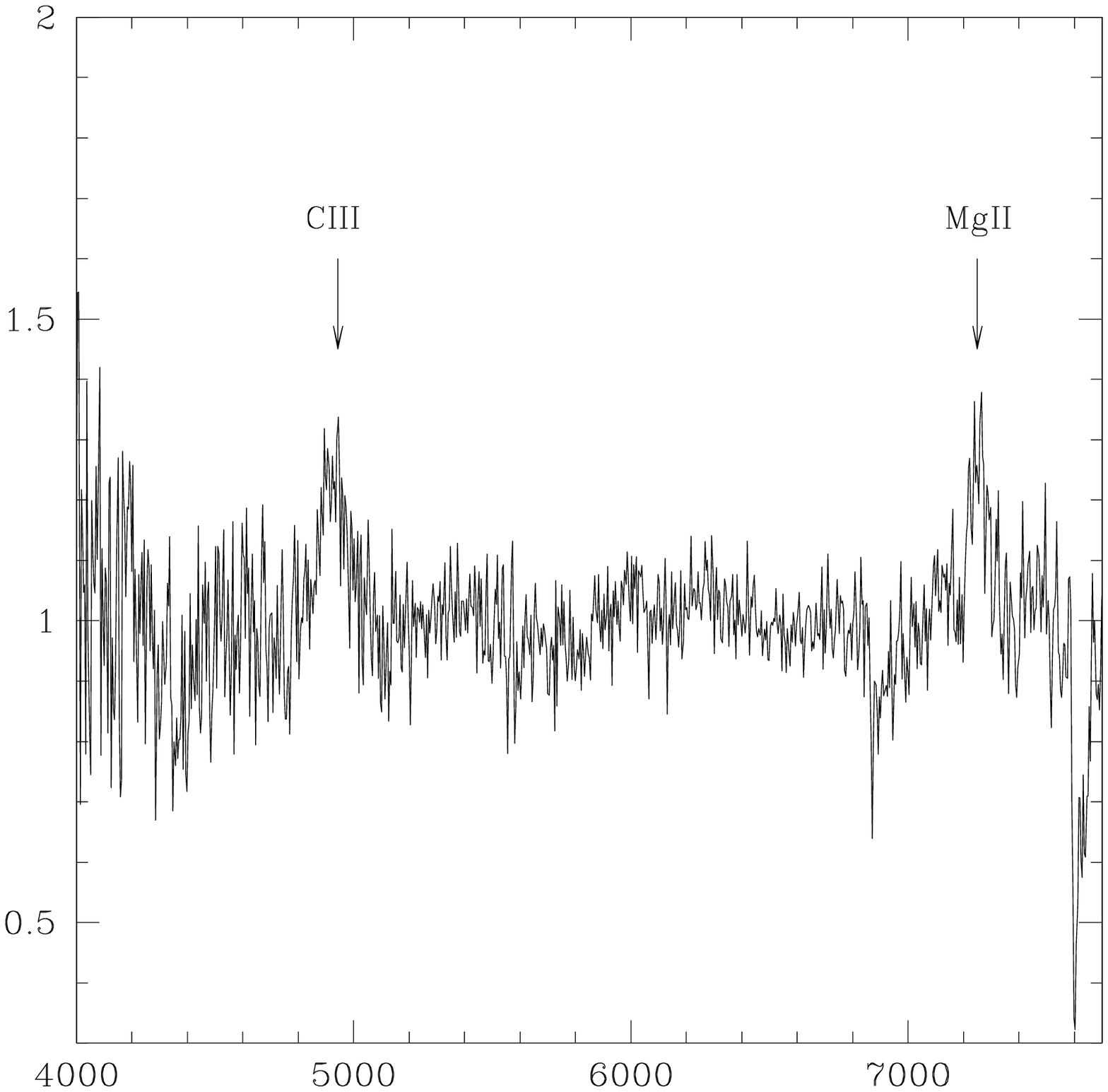}{$\lambda$, $\AA$}{$F_\lambda$, relative intensity}

\hspace{-2mm}\includegraphics[width=7.9cm,bb=15 165 565 695]{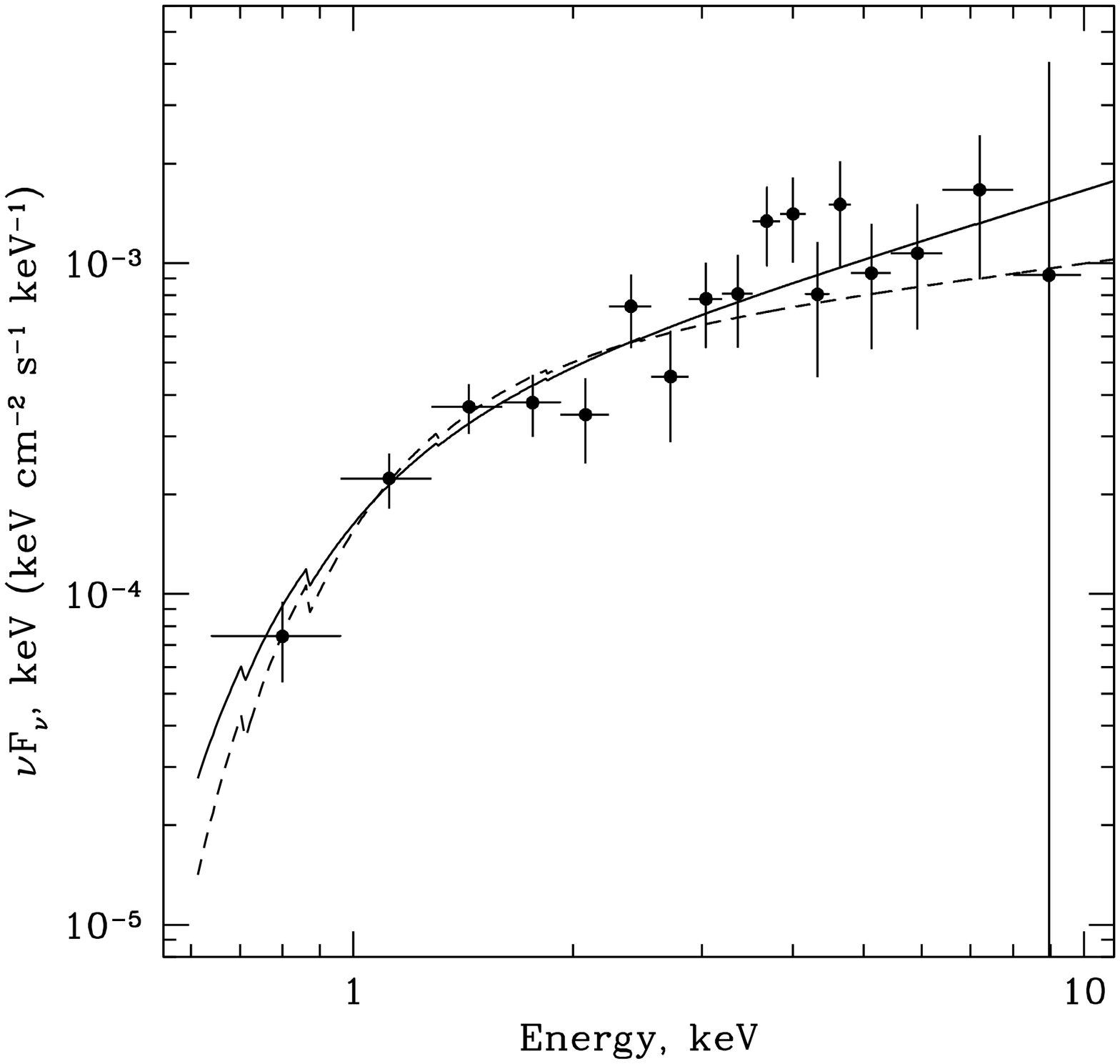}
}

\caption{\small (a) Sky field around \emph{SWIFT J0113.8+2515} from SDSS data; the contours indicate the soft X-ray intensity
levels from XRT data, the arrow indicates of the optical counterpart. (b) The optical spectrum of \emph{SWIFT J0113.8+2515}
taken with the BTA telescope. The two emission lines we managed to identify are labeled. (c) The energy spectrum of
\emph{SWIFT J0113.8+2515} from XRT data: the solid and dashed lines indicate, respectively, the best fits (see the text)
by a power law and a power law with a slope of 1.7 and absorption.}\label{swift01138}

\end{figure*}

As we see from Fig.\ref{swift01138}a, one object with $m_r=19.1$ (from SDSS data) falls into the XRT error circle
of the source (RA=01$^h$ 13$^m$ 22.7$^s$, Dec=25\fdg 18\arcmin 54\arcsec, J2000, the localization accuracy is $\sim4$\arcsec). The optical
spectrum of this object taken with the BTA telescope reveals two relatively weak emission lines (Fig.\ref{swift01138}b).
We identified these lines with the broad CIII $1909\AA$ and MgII $2798\AA$ lines with a Doppler
width of $\sim5000$ km/s ($FWHM$) at redshift $z=1.594\pm0.002$. Thus, our measurements indicate that
\emph{SWIFT J0113.8+2515} is a quasar.

\begin{figure}
\centering
\includegraphics[width=8.8cm,bb=30 175 550 660]{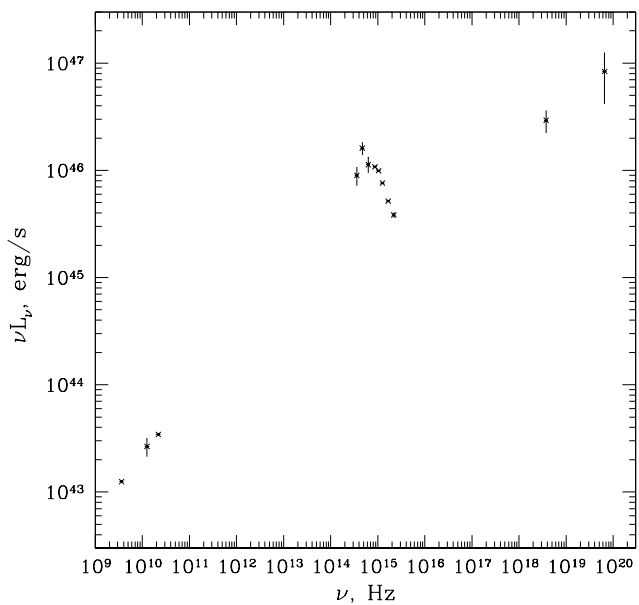}

\caption{\small Spectral energy distribution for \emph{SWIFT J0113.8+2515} based on data from different radio telescopes, the infrared
2MASS survey, the optical SDSS survey, the XRT telescope, and the hard X-ray BAT instrument (both onboard the SWIFT
observatory). The infrared and optical measurements were corrected for Galactic extinction with $E(B-V) = 0.080$ (Schlegel
et al. 1998). }\label{spectr01138}

\end{figure}

Our search in astronomical catalogs showed that apart from the optical band, the object was previously
detected in the near infrared (in the 2MASS twomicron survey) and with several radio telescopes. In
addition, it was detected at a low confidence level in the ROSAT all-sky X-ray survey (the source
\emph{RX J0113.2+2518}; Brinkmann et al. 1997). Figure \ref{spectr01138} shows the source's spectral energy distribution from
the available data. We calculated the luminosities using the cosmological model with $\Omega_{\rm m}=0.3$, $\Omega_\Lambda=0.7$ and
$H_0=72$~km/s/Mpc. The flat (with a spectral slope $\alpha\sim 0$) radio spectrum and the presence of two
emission peaks (in the optical band and in the hard X-ray or harder energy band) indicate that the object
is a flat-spectrum radio quasar, i.e. a blazar. The derived spectral energy distribution is in satisfactory
agreement with the well-known blazar sequence: the dependence of the positions of the two (synchrotron
and Comptonization) emission peaks on the object's luminosity (e.g., Fossati et al. 1998).

The soft-X-ray spectrum of the source is well fitted by a simple power law, possibly modified by absorption
(Fig.\ref{swift01138}c). Assuming that the absorption occurs inside or near the source, i.e. at redshift $z=1.594$,
the best values of the photon index and the column density of neutral matter turn out to be
$\Gamma=1.10\pm0.20$, $N_H=2.3^{+1.8}_{-1.0}\times10^{22}$ cm$^{-2}$. The $\chi^2$ value
per degree of freedom is 0.73(26). The derived upper limit for the absorbtion column density turns out to be
considerably higher (even if the redshift is taken into account) than the column density of photoabsorbing
matter on the line of sight in the Galaxy, $N_H=0.06\times10^{22}$ cm$^{-2}$ (Dickey and Lockman 1990). The BAT
measurement of the flux in the harder $14-195$ keV energy band confirms that the hard spectrum typical
of blazars (flat-spectrum radio quasars) extends at least to the range of soft gamma-ray energies. The
power-law model with a fixed slope $\Gamma=1.7$ and absorption $N_H=(5\pm 1)\times10^{22}$ cm$^{-2}$
at $z=1.594$ that could correspond to an ordinary quasar (not of the blazar type) describes the X-ray spectrum slightly
more poorly ($\chi^2=0.95$ per degree of freedom).

The flux from the source in the $2-10$ keV energy band, $F_{X}\simeq3\times10^{-12}$ erg cm$^{-2}$ s$^{-1}$, corresponds
to a luminosity $L_{X}\simeq1.9\times10^{46}$ erg s$^{-1}$ (without a correction for the absorption). Thus,
\emph{SWIFT J0113.8+2515} is one of the brightest blazars detected in the INTEGRAL and Swift all-sky surveys.

\bigskip

{\bf SWIFT J2237.2+6324.} Three soft X-ray sources were detected in the XRT error circle of
\emph{SWIFT J2237.2+6324}. The emission from two of them turns out to be considerably less intense
than that from the third object with coordinates RA=22$^h$ 36$^m$ 37.3$^s$, Dec=63\fdg 29\arcmin 31\arcsec (J2000)
and a localization accuracy of $\sim4$\arcsec. The latter circumstance makes this object the most probable source of the
observed emission in the hard X-ray energy band as well. The sky field containing \emph{SWIFT J2237.2+6324}
was observed with the XRT telescope three times in 2008 (on July 9, 16, and 17) with a total exposure
time of $\simeq16$ ks. During these observations, the flux from the program source changed by a factor of $\approx2$.
However, we found no ensuing significant changes in the shape of the spectrum, possibly because the
quality of the latter was not very high, which, in turn, is related to the faintness of the object itself. Thus,
all of the subsequent reasoning refers to the total spectrum obtained during the above observations.

\begin{figure}
\centering
\vbox{

~\hspace{5mm}\includegraphics[width=7.0cm,bb=40 140 575 635]{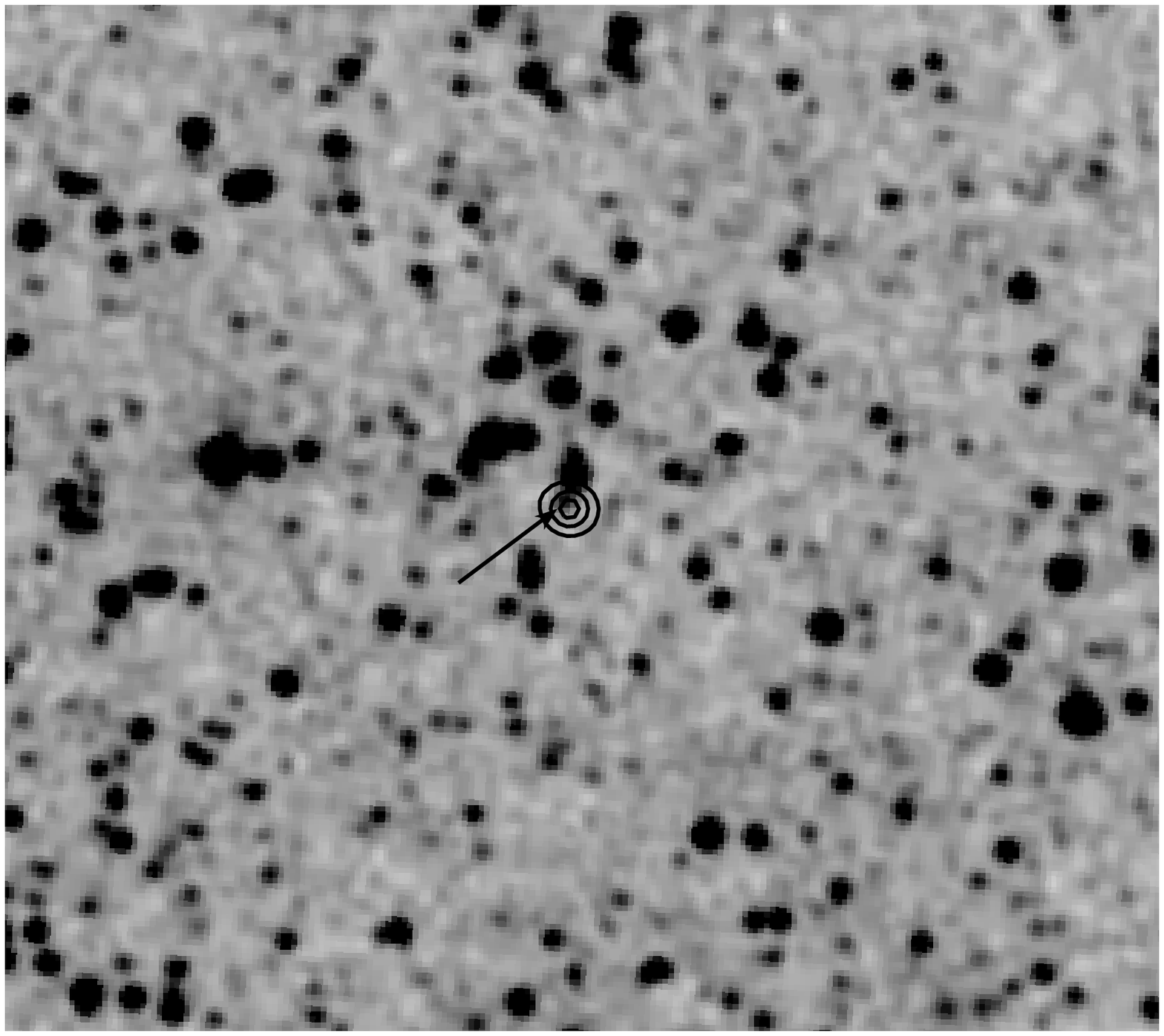}
\vspace{0mm}\smfigure{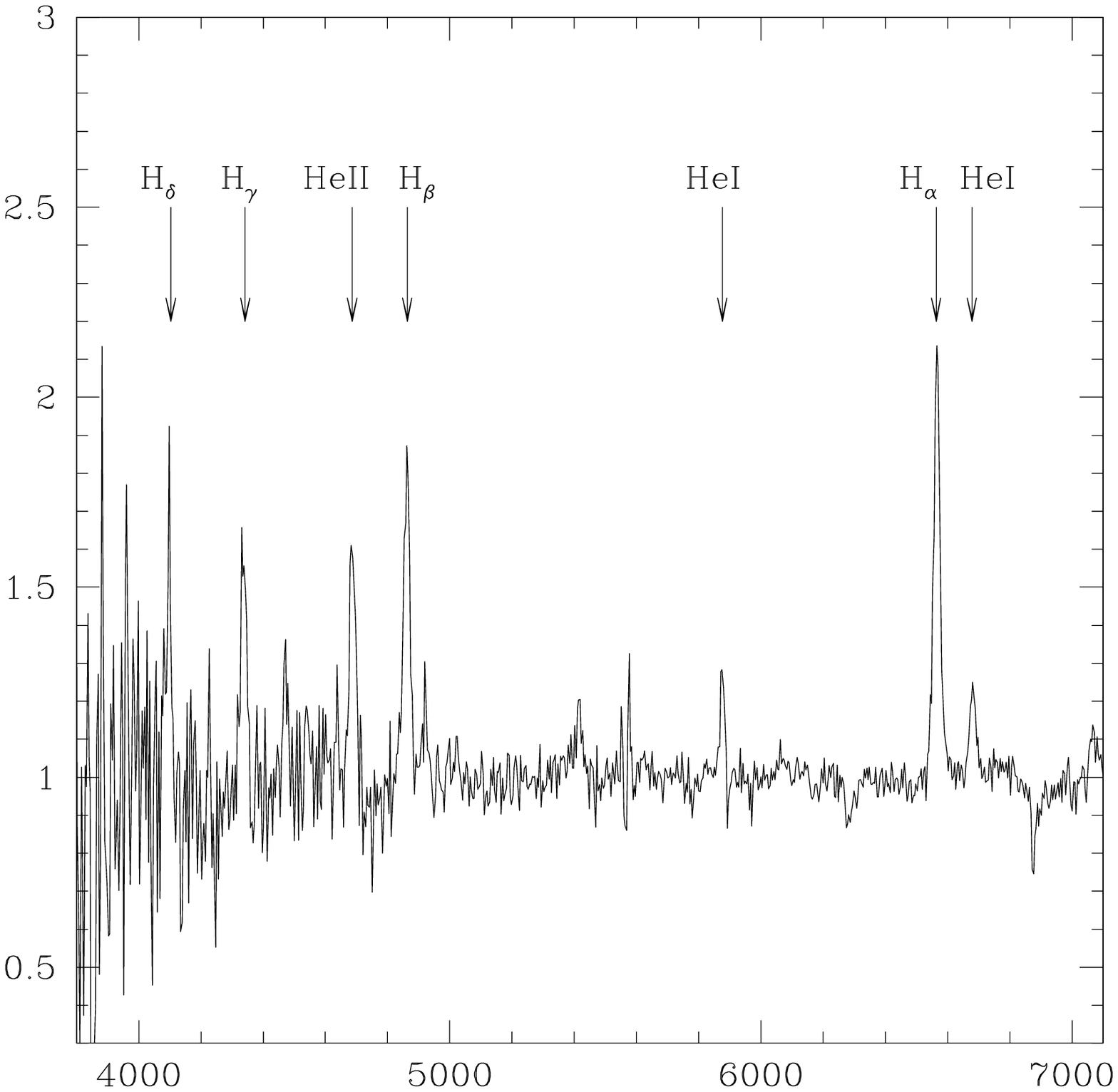}{$\lambda$, $\AA$}{$F_\lambda$, relative intensity}
\hspace{-3mm}\includegraphics[width=7.9cm,bb=15 165 565 695]{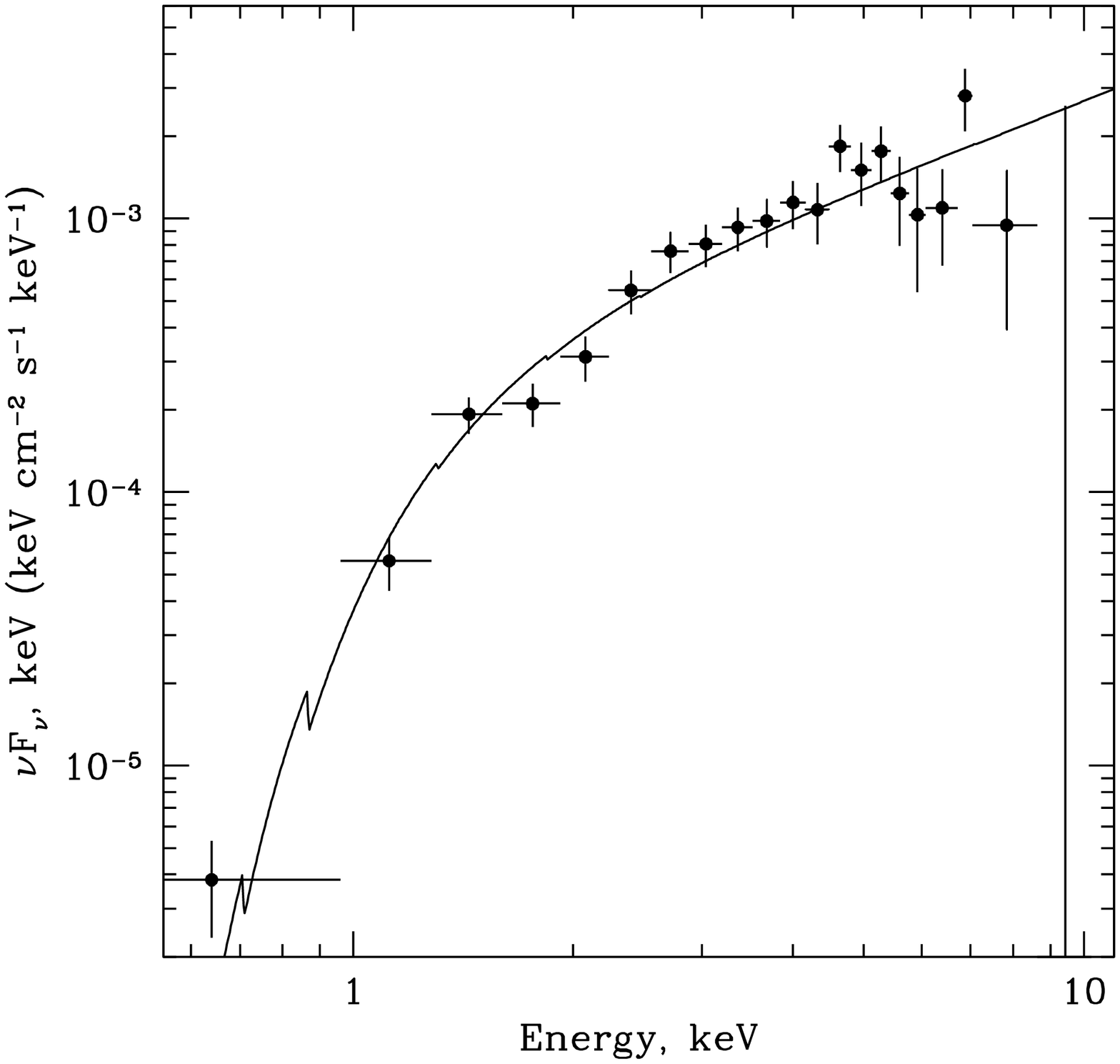}
}

\caption{\small (a) Sky field near \emph{SWIFT J2237.2+6324} from digitized POSS) plates; the contours indicate the soft X-ray intensity
levels from XRT data; the arrow indicates the position of the optical counterpart. (b) The optical spectrum of
\emph{SWIFT J2237.2+6324} taken with the BTA telescope. The main emission lines are labeled. (c) The energy
spectrum of \emph{SWIFT J2237.2+6324} from XRT data: the solid line indicates the best fit.}\label{swift2237}

\end{figure}

Comparison of the soft X-ray images with the digitized Palomar Observatory Sky Survey (POSS) plates indicates that there is a
bright optical object near the improved error circle of \emph{SWIFT J2237.2+6324}, but it does not fall into
this circle (see Fig.\ref{swift2237}a). The most probable optical counterpart in the system is the fainter star with $m_I\simeq18$
lying several arcseconds southwest of the brighter star (indicated by the arrow in Fig.\ref{swift2237}a). To establish
its nature, we performed spectroscopic observations with the BTA telescope; during these observations,
the slit was positioned in such a way that both objects fell into it simultaneously. The spectrum of
the optical counterpart of \emph{SWIFT J2237.2+6324} is shown in Fig.\ref{swift2237}b. It exhibits a set of bright
hydrogen and helium emission lines characteristic of the spectra of accretion disks around white dwarfs
(see, e.g., Williams and Ferguson 1982). Thus, this object is a cataclysmic variable, an accreting
white dwarf in a binary system. It is important to note that the width of the lines detected in the
spectrum($FWHM \simeq15-20\AA$) exceeds considerably the spectral resolution of the grating used ($\approx6\AA$),
suggesting their internal broadening. Such a broadening can be due to the rotation of emitting matter
in the accretion disk around the white dwarf. The velocities derived from the observed emission line
widths ($v\simeq300-450$ km/s) are in agreement with the direct measurements in similar systems (e.g.,
Szkody et al. 2001).

The soft X-ray spectrum of \emph{SWIFT J2237.2+6324} can be fitted by a power-law dependence of the
photon flux density on energy modified by photoabsorption with the following parameters (Fig.\ref{swift2237}c):
$\Gamma=0.95\pm0.17$, $N_H=(0.77\pm0.17)\times10^{22}$ cm$^{-2}$, the flux in the $2-10$ keV energy band
$F_{X}=3.5^{+0.4}_{-0.8}\times10^{-12}$ erg cm$^{-2}$ s$^{-1}$. Note that the derived absorption
agrees well with the interstellar absorption toward the source, $N_H=0.76\times10^{22}$ atoms cm$^{-2}$ (Dickey and
Lockman 1990).

\bigskip

\begin{figure*}
\centering

\vbox{
~\hspace{5mm}\includegraphics[width=7.2cm,bb=40 140 575 635]{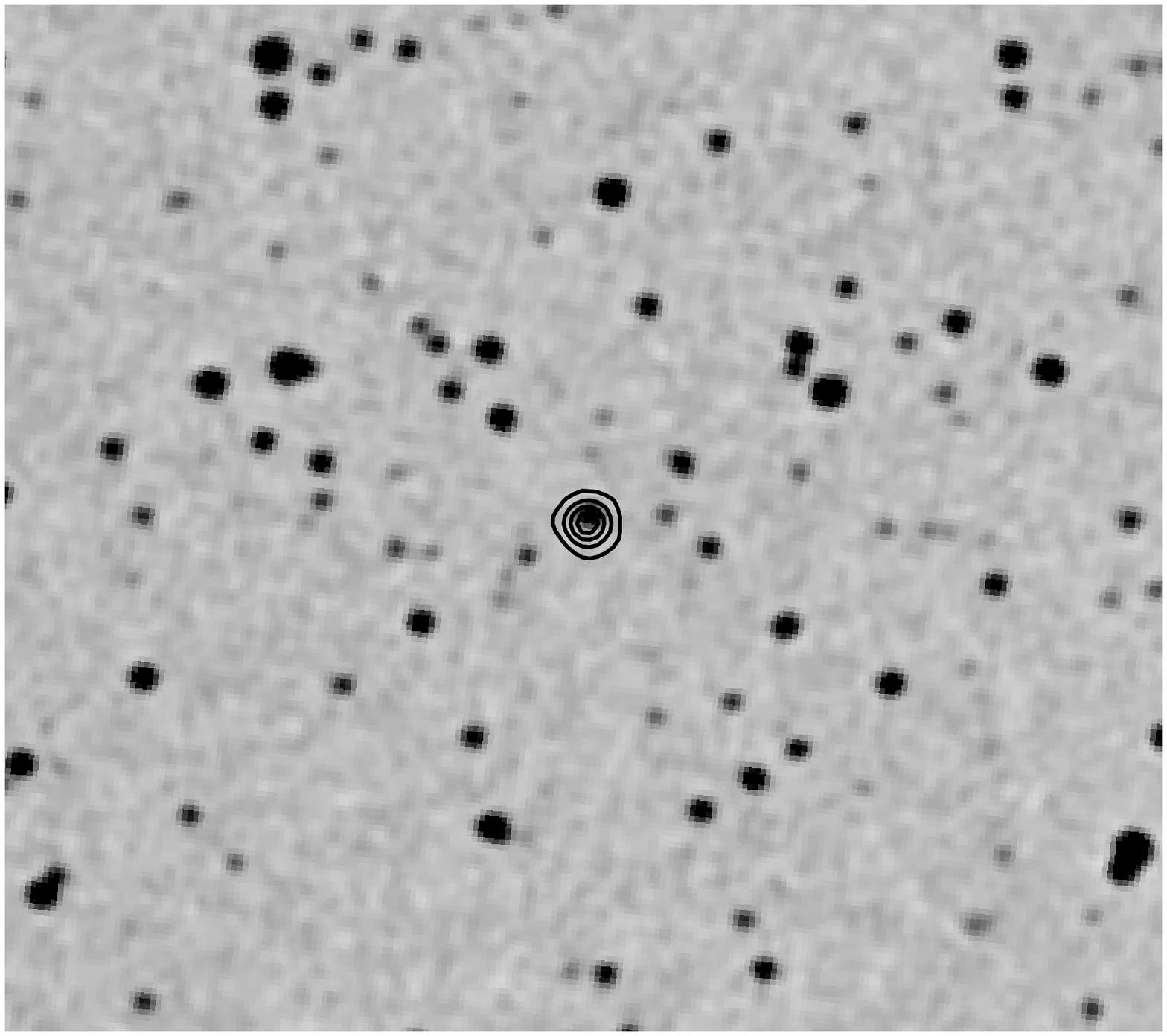}

\smfigure{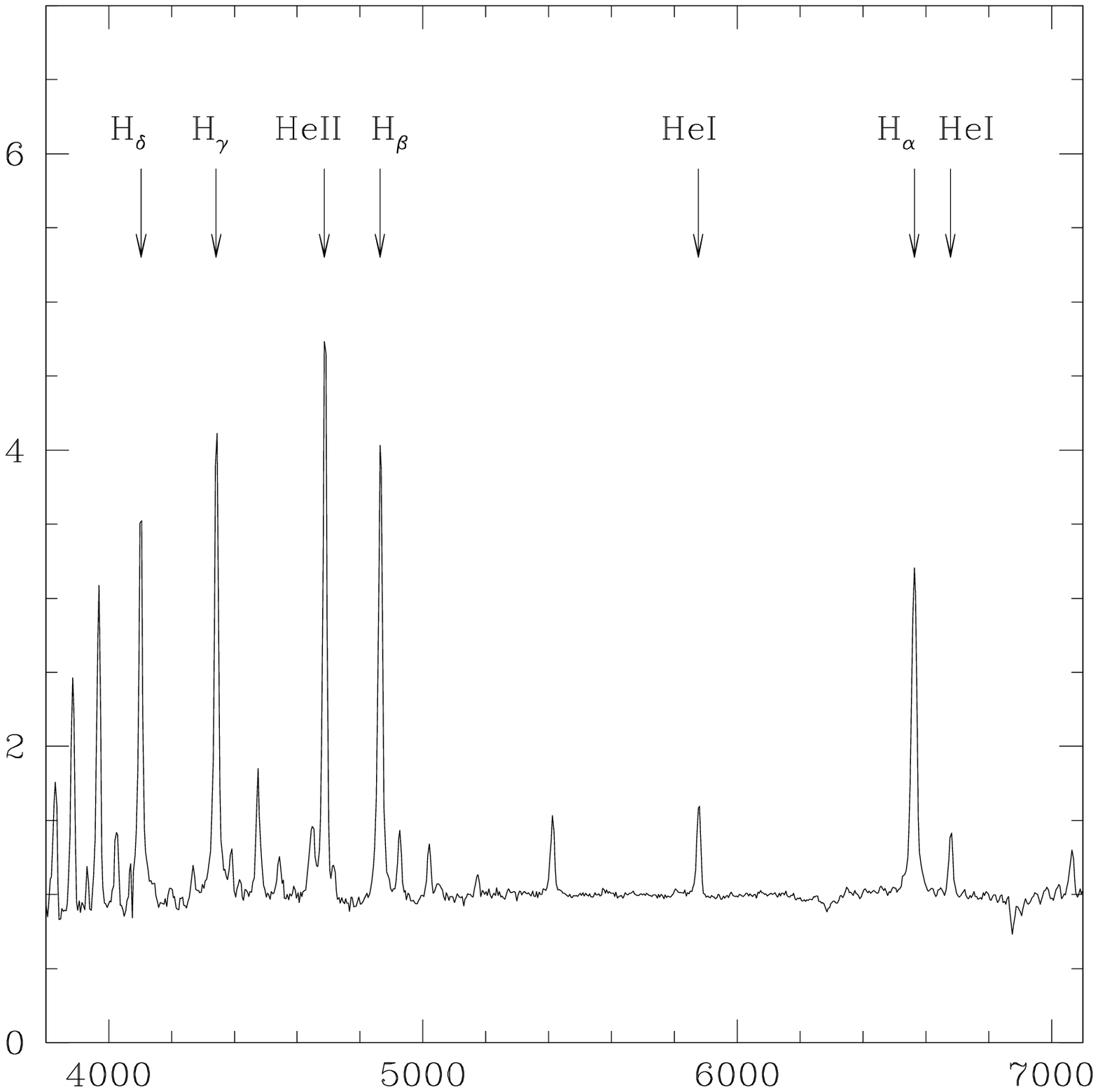}{$\lambda$, $\AA$}{$F_\lambda$, relative intensity}

\hspace{3mm}\includegraphics[width=8.5cm,bb=50 180 590 465]{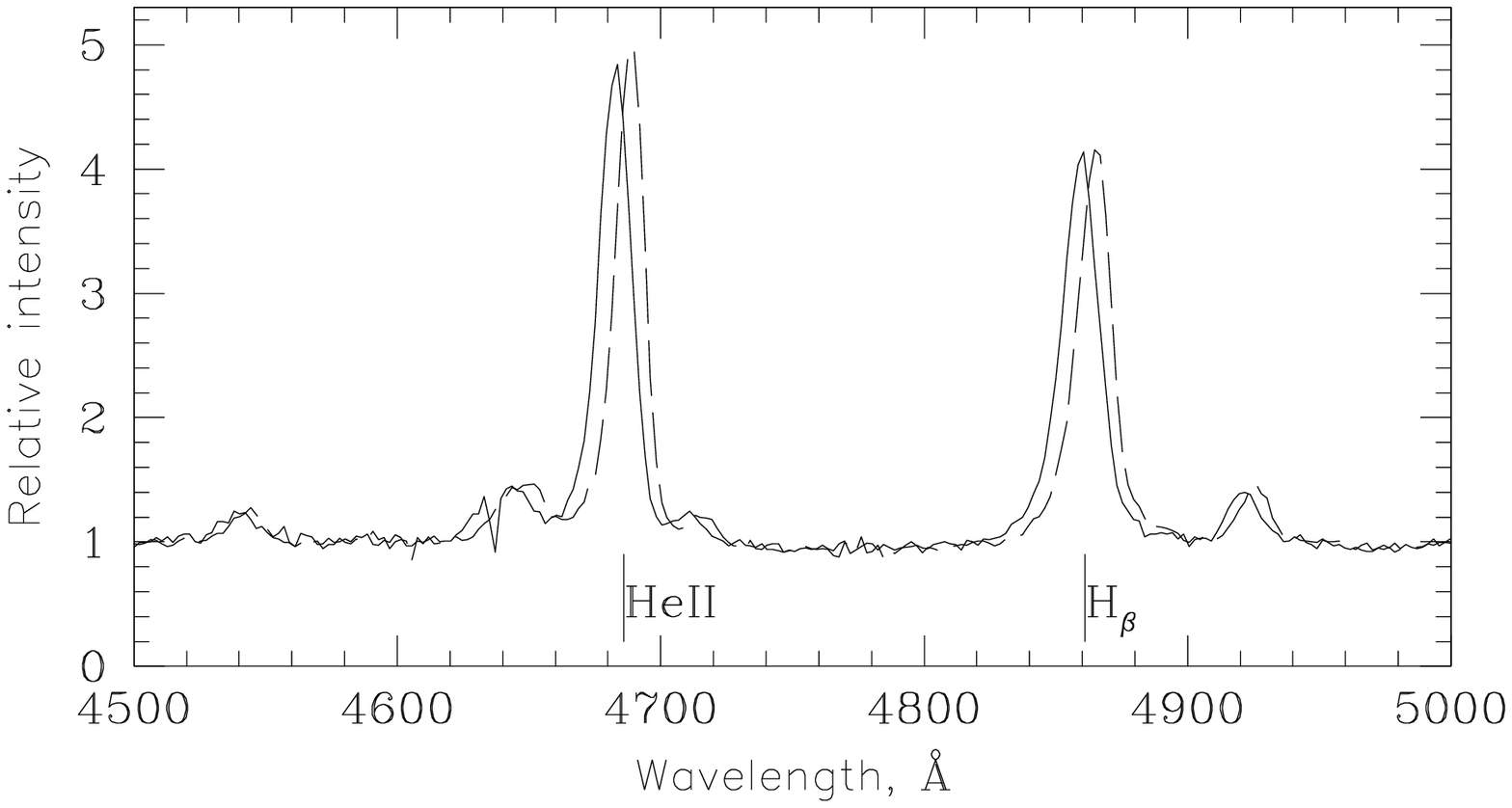}
}

\caption{\small (a) Image of the sky field around \emph{SWIFT J2341.0+7645} from digitized POSS plates; the contours indicate the soft
X-ray intensity levels from XRT data. (b) The first of the two optical spectra for \emph{SWIFT J2341.0+7645} taken with the BTA
telescope. (c) A magnified image of the spectral region in the wavelength range $4500-5000\AA$. The solid line indicates the first
spectrum of the source and the dashed line indicates the second spectrum taken $\approx18$ min after the first one. A general shift of
the second spectrum relative to the first one toward lower frequencies is clearly seen.}\label{swift23410}

\end{figure*}

{\bf SWIFT J2341.0+7645.} The XRT observations on July 31 and August 4, 2010, with a total exposure
time of $\simeq10.5$ ks allowed \emph{SWIFT J2341.0+7645} to be unambiguously identified with a soft X-ray
source with coordinates (J2000) RA=23$^h$ 40$^m$ 20.6$^s$, Dec=76\fdg 42\arcmin 09\arcsec and a localization accuracy of
$\sim3$\arcsec (the latter coincides with the X-ray source \emph{1RXS J234015.8+764207} from the ROSAT survey).
Comparison of the X-ray and optical images shows that only one optical object with $m_I \simeq17$ falls into
this error circle (Fig.\ref{swift23410}a).

The optical spectrum of the source exhibits a large set of hydrogen and helium lines (at redshift
$z=0$) typical of the spectra of accreting white dwarfs (Fig.\ref{swift23410}b). The emission line width ($20-25 \AA$) clearly
points to the line formation in an accretion disk around a compact object and corresponds to matter
velocities in the disk $v\simeq500-800$ km/s, which are significantly higher than those in \emph{SWIFT J2341.0+7645}
but, nevertheless, are also observed in systems with white dwarfs (Howell et al. 2003). During one night,
we took two spectra of the source with an interval of about 18 min. We found that the positions of the
emission lines in these spectra shifted by $4-5 \AA$, which exceeds considerably the positional error of
the line centroids (Fig.\ref{swift23410}c). A natural explanation of the observed effect is the orbital motion in the
binary, but additional photometric and spectroscopic observations are required for firm conclusions to be
reached and for the orbital period to be measured. It should be noted that the shift of various emission
lines is inconsistent with the simple model in which all lines originate in the same part of the binary system.
Such a behavior is not unusual, because different parts of the binary system (moving with different
velocities around the system's center of mass) contain different sets of emission lines (see, e.g., the reviews
on Doppler tomography of binary systems by Marsh and Horne (1988) and Steeghs et al. (1997)).

\begin{figure}
\centering

\includegraphics[width=8.5cm,bb=15 165 565 700]{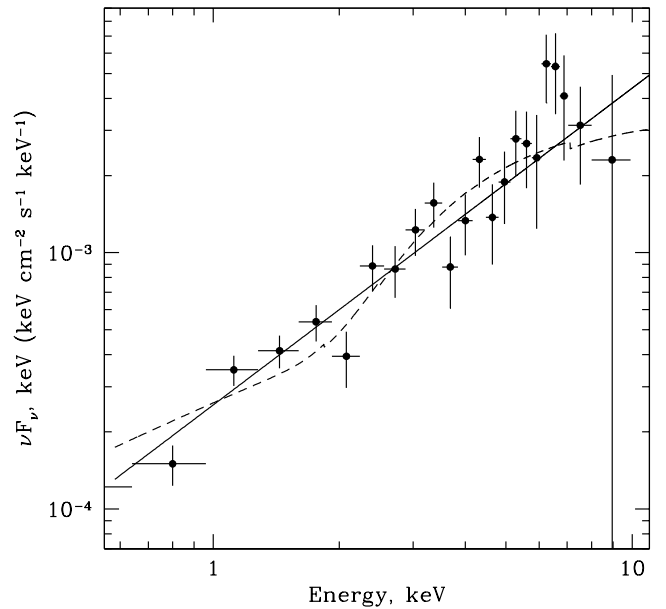}

\caption{\small XRT X-ray spectrum of \emph{SWIFT J2341.0+7645}. The solid line indicates the best fit; the dashed line
indicates the fit by the model of partially absorbed bremsstrahlung with a temperature of 20 keV.}
\label{spectr_23410}

\end{figure}

Owing to the relatively long XRT observation, we managed to obtain a high-quality energy spectrum
of the source (Fig.\ref{spectr_23410}). The spectrum can be satisfactorily described by a power law with a
slope $\Gamma=0.71\pm0.08$ and a flux $F_{X}=(5.3\pm0.4)\times10^{-12}$ erg cm$^{-2}$ s$^{-1}$ in the $2-10$
keV energy band. It is interesting to note that at energies $\sim6.5$ keV, there is evidence for the presence of a set of emission lines
($6.4, 6.7, 6.9$ keV) typical of accreting white dwarfs. Adding a Gaussian at these energies to the model
formally improves the quality of the fit to the spectrum but does not allow the line parameters to be properly
constrained: the confidence level of the line is about $3\sigma$ and its equivalent width is $\sim0.8$ keV.

In conclusion, note that fitting the spectrum by the model of partially absorbed bremsstrahlung with a
temperature of 20 keV (characteristic of white dwarfs in binary systems; see, e.g., Suleimanov et al. 2005)
gives a poorer $\chi^2$ value, but the model parameters agree satisfactorily with those observed in similar
systems: the absorption column density is $N_H\simeq6\times10^{22}$ atoms cm$^{-2}$ and the fraction of the absorbed emission
in neutral matter is $C_{F}\simeq0.73$ (Fig.\ref{spectr_23410}, the dashed line).

Summarizing the aforesaid, we can assert that \emph{SWIFT J2341.0+7645} is a cataclysmic variable, an
accreting white dwarf in a binary system.

\section{Conclusions}
\label{sec:concl}

We performed optical identifications of four hard X-ray sources from the Swift all-sky survey. Two
sources were shown to be extragalactic in nature: \emph{SWIFT J0003.3+2737} is a Seyfert 2 galaxy at redshift
$z=0.03946\pm0.00002$ and \emph{SWIFT J0113.8+2515} is a blazar with a luminosity $L_{X}\simeq1.9\times10^{46}$ erg s$^{-1}$ at redshift
$z=1.594$. The other two objects, \emph{SWIFT J2237.2+6324} and \emph{SWIFT J2341.0+7645}, belong to the class of cataclysmic variables (accreting
white dwarfs) in our Galaxy. For \emph{SWIFT J2341.0+7645}, we detected a spectral line shift during the observations,
which is most likely due to the orbital motion in the binary system.

\bigskip
~\bigskip

\acknowledgements

This study was supported by the Russian Foundation for Basic Research (project nos. 10-02-00492,
10-02-01442, 10-02-00463, 11-02-12285-ofi-m-2011, 11-02-12271-ofi-m-2011), the Presidium of Russian Academy of Sciences (the
''Non-stationary phenomena in objects of the Universe'' and ''Active and Stochastic Processes in the Universe''
Programs), the Program of the the President of the Russian Federation for support of scientific
schools (grant NSh-5603.2012.2), grant no. MD-1832.2011.2 from the President of the Russian Federation,
grants from the Dynasty Foundation, and the State Contract no. 14.740.11.0611. We are grateful
to S. Fabrika (Special Astrophysical Observatory, Russian Academy of Sciences) for his help in organizing
the observations with the BTA telescope and for a discussion of the results.

\parindent=0mm
1. V. L. Afanasiev and A. V. Moiseev, Astron. Lett. 31, 194 (2005).

2. E. Barlow, C. Knigge, A. Bird, et al., Mon. Not. R. Astron. Soc. 372, 224 (2006).

3. W. Baumgartner, J. Tueller, C. Markwardt, G. Skinner, Bulletin of the American Astronomical Society, 41, 675 (2010)

4. I. Bikmaev, M. Revnivtsev, R. Burenin, and R. Syunyaev, Astron. Lett. 32, 588 (2006).

5. I. Bikmaev, R. Burenin, M. Revnivtsev, et al., Astron. Lett. 34, 653 (2008).

6. A. Bird, A. Bazzano, L. Bazzani, et al., Astrophys. J. Suppl. Ser. 186, 1 (2010).

7. R. Burenin, A. Meshcheryakov, M. Revnivtsev, et al., Astron. Lett. 34, 367 (2008).

8. R. Burenin, I. Bikmaev, M. Revnivtsev, et al., Astron. Lett. 35, 71 (2009).

9. G. Cusumano, V. La Parola, A. Segreto, et al., Astron. Astrophys. 524, 64 (2010).

10. J. Dickey and F. Lockman, Ann. Rev. Astron. Astrophys. 28, 215 (1990).

11. E. Filippova, S. Tsygankov, A. Lutovinov, and R. Syunyaev, Astron. Lett. 31, 729 (2005).

12. G. Fossati, L. Maraschi, A. Celotti, et al., Mon. Not. R. Astron. Soc. 299, 433 (1998).

13. N. Gehrels, G. Chinkarini, P. Giommi, et al., Astrophys. J. 611, 1005 (2004).

14. S. Howell, A. Adamson, and D. Steeghs, Astron. Astrophys. 399, 219 (2003).

15. R. Krivonos, M. Revnivtsev, A. Lutovinov, et al., Astron. Astrophys. 475, 775 (2007).

16. R. Krivonos, S. Tsygankov, M. Revnivtsev, et al., Astron. Astrophys. 523, A61 (2010).

17. A. Lutovinov, M. Revnivtsev, M. Gilfanov, et al., Astron. Astrophys. 444, 821 (2005).

18. A. Lutovinov and S. Tsygankov, Astron. Lett. 35, 433 (2009).

19. A. Lutovinov, R. Burenin, M. Revnivtsev, et al., Astron. Lett. 36, 904 (2010).

20. A. Lutovinov, R. Burenin, M. Revnivtsev, and I. Bikmaev, Astron. Lett. 38, 1 (2012).

21. T. Marsh and K. Horne, Mon. Not. R. Astron. Soc. 235, 269 (1988).

22. N. Masetti, R. Landi, M. Pretorius, et al., Astron. Astrophys. 470, 331 (2007).

23. N. Masetti, P. Parisi, E. Palazzi, et al., Astron. Astrophys. 519, 96 (2010).

24. M. Revnivtsev, A. Lutovinov, E. Churazov, et al., Astron. Astrophys. 491, 209 (2008).

25. S. Sazonov, E. Churazov, M. Revnivtsev M., et al., Astron. Astrophys. 444, L37 (2005)

26. S. Sazonov, M. Revnivtsev, R. Krivonos, et al., Astron. Astrophys. 462, 57 (2007).

27. S. Sazonov, R. Krivonos, M. Revnivtsev, et al., Astron. Astrophys. 482, 517 (2008)

28. D. J. Schlegel, D. P. Finkbeiner, and M. Davis, Astrophys. J. 500, 525 (1998).

29. D. Steeghs, E. Harlaftis, and K. Horne, Mon. Not. R. Astron. Soc. 290, L28 (1997).

30. V. Suleimanov, M. Revnivtsev, and H. Ritter, Astron. Astrophys. 435, 191 (2005).

31. P. Szkody, K. Nishikida, K. Long, and R. Fried, Astron. J. 121, 2761 (2001).

32. J. Tomsick, S. Chaty, J. Rodriguez, et al., Astrophys. J. 701, 811 (2009).

33. J. Tueller, W. Baumgartner, C. Markwardt, et al., Astrophys. J. Suppl. Ser. 186, 378 (2010).

34. R. E. Williams and D. H. Ferguson, Astrophys. J. 257, 672 (1982).

35. C. Winkler, T. Courvoisier, G. Di Cocco, et al., Astron. Astrophys. 411, L1 (2003).

\end{document}